# SPATIAL PROCESSING FRONT-END FOR DISTANT ASR EXPLOITING SELF-ATTENTION CHANNEL COMBINATOR

*Dushyant Sharma[1], Rong Gong[2], James Fosburgh[1], Stanislav Yu. Kruchinin[2],*
*Patrick A. Naylor[3] and Ljubomir Milanović[2]*

[1]Research & Development, Nuance Communications Inc., USA
[2]Research & Development, Nuance Communications GmbH, Austria
[3]Electronic & Electrical Engineering, Imperial College London, UK

## ABSTRACT

We present a novel multi-channel front-end based on channel shortening with the Weighted Prediction Error (WPE) method followed by a fixed MVDR beamformer used in combination with a recently proposed self-attention-based channel combination (SACC) scheme, for tackling the distant ASR problem. We show that the proposed system used as part of a ContextNet based end-to-end (E2E) ASR system outperforms leading ASR systems as demonstrated by a 21.6% reduction in relative WER on a multi-channel LibriSpeech playback dataset. We also show how dereverberation prior to beamforming is beneficial and compare the WPE method with a modified neural channel shortening approach. An analysis of the non-intrusive estimate of the signal C50 confirms that the 8 channel WPE method provides significant dereverberation of the signals (13.6 dB improvement). We also show how the weights of the SACC system allow the extraction of accurate spatial information which can be beneficial for other speech processing applications like diarization.

## 1. INTRODUCTION

The problem of Distant Automatic Speech Recognition (DASR) remains an important topic that has received much attention in the past few years. The current state of the art methods for DASR are based on multi-channel end-to-end (MCE2E) architectures where the front-end signal processing, acoustic modeling and language modeling are integrated into a single system that allows for a joint optimization of all sub-systems [1–4].

The current state of the art MCE2E DASR systems can be broadly grouped into three categories. The first are systems where the multi-channel front-end is based on a signal processing based spatial filtering method with some elements of the algorithm being jointly estimated with the E2E ASR system [5]. In [6] the authors use neural networks to estimate spectral masks from which the cross-power spectral density matrices of speech and noise are estimated and used to compute the beamformer coefficients. Furthermore, the authors show how the trained network is independent of microphone array configuration and the proposed method provides perceptual quality and ASR gains. A similar approach was proposed in [7] where the authors also use a neural mask estimation based MVDR beamforming front end, with a reference microphone selection via an attention network as part of a CTC and attention hybrid sequence-to-sequence ASR system.

The second category of systems are based on neural filter-and-sum structures that directly optimize the ASR loss function. In [8] the authors propose the neural adaptive beamformer where raw time-domain microphone signals are input to a Long Short Term Memory (LSTM) neural network, the outputs of which are the filter coefficients that are used by a filter-and-sum operation to generate the input signal to the E2E ASR system. In [4] the authors propose a time invariant filter-and-sum structure where a set of filters are learnt during training, optimizing the ASR loss function and then fixed during the test phase. This is claimed to give a robust solution that outperforms classical beamforming approaches. Finally there are methods for speech dereverberation, which target the reduction of the level of reverberation in the signals. The complete inversion of the acoustic transfer functions is a challenging problem and thus most state of the art methods target impulse response shortening, whereby some amount of early reflections are kept intact and only the later reverberation is removed, making the problem more feasible. This also has the advantage that the spatial localization information remains intact and allows other spatial processing algorithms to be applied as subsequent processing. The Weighed Prediction Error (WPE) [9] algorithm is an important method in this category, where the reverberant tail is estimated and removed from the input signal with the constraint of retaining some amount of early reflections (which have been shown to be beneficial for speech reception), thus improving the C50 metric. Another recent approach is based on neural channel shortening that used a U-Net encoder-decoder architecture to estimate a channel shortening filter, which when convolved with the input signal achieves dereverberation [10]. In this paper, we consider a typical meeting room scenario where a wall-mounted microphone array is used to capture the interaction between participants conversing in the room. We consider the scenario where the speakers are engaged in conversations and are positioned at different distances, azimuths and orientations relative to the array (this is in contrast to a meeting room scenario with a microphone array placed in the centre of the table with participants seated around, facing the array). The biggest challenge in such a scenario for ASR is the adverse effect of room reverberation. We propose a novel front-end based on a dereverberation of the microphone signals by a multi-channel Weighted Prediction Error (WPE) based method [9] followed by a fixed beamformer with 16 beams driving a recently proposed front-end for multi-channel signal combination using a self-attention mechanism [11].

## 2. METHODS

In the following description of the spatial processing front-end, we assume that an $M$-dimensional signal is captured using a microphone array with $M$ microphones and represented as the vector $\mathbf{y}_{t,f}$ in the short-time Fourier transform (STFT) domain, with $t$ denoting the time frame index and $f$ the frequency bin index (we omit the microphone index here for brevity). The signal model we adopt is based on impulse response (or channel) shortening, where the room impulse response (RIR) is decomposed into an early and late part and thus the recorded signal at microphone $m$ can be expressed as follows, $\mathbf{y}_{t,f} = \mathbf{x}_{t,f}^{(\text{early})} + \mathbf{x}_{t,f}^{(\text{tail})}$, where $\mathbf{x}_{t,f}^{(\text{early})}$ and $\mathbf{x}_{t,f}^{(\text{tail})}$ are the STFTs of the source signal which is convolved with the early part of the RIR and with the late reflections, respectively. With this signal model, it is possible to apply channel shortening as a pre-process to a beamforming method which combines the $M$ signals into a set of $N_b$ single-channel beams. These can then be used as input to an E2E ASR system.

### 2.1. Weighted Prediction Error (WPE)

The WPE algorithm estimates the reverberation tail of the signal and subtracts it from the observation to obtain an estimate of the direct signal component and early reflections [9]. In our simulations, we use the multiple-input multiple-output (MIMO) and single-input single-output (SISO) versions of WPE [12] implemented in the NaraWPE code [13]. Given the filter weights $\mathbf{G}_f \in \mathbb{C}^{MN \times M}$, an estimate of the clean speech with early reflections is obtained as $\hat{x}_{t,f,m}^{(\text{early})} = y_{t,f,m} - \sum_{\tau=\Delta}^{\Delta+N-1} \sum_{m'=1}^{M} g_{\tau,f,m,m'}^* y_{t-\tau,f,m'}$. A delay term $\Delta > 0$ is introduced to avoid whitening of the speech source, $N$ is the number of filter taps, $m$ is the microphone index. WPE maximizes the likelihood of the model under assumption that the single-channel direct signal is a realization of a zero-mean complex Gaussian with the channel-independent variance estimated iteratively with the filter weights [13].

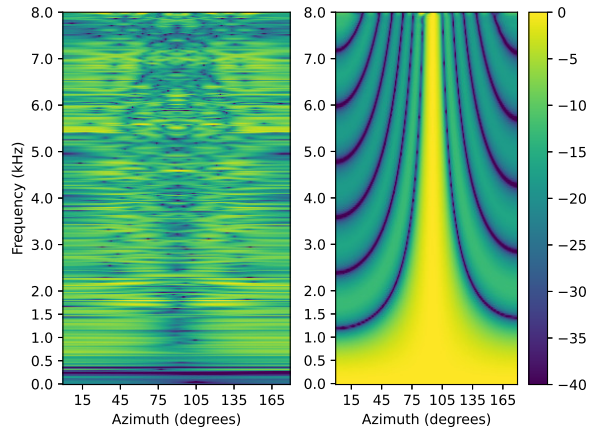

**Fig. 1**. Beampatterns for the 9$^{\text{th}}$ filter of the trained NSF beamformer (left panel) and for the FMBu beamformer in a 95° look direction (right panel). The colorbar shows the gain (dB) for a particular azimuth and frequency.

### 2.2. Channel Shortening (CS)

A recent, deep learning based approach for channel shortening using the U-Net encoder-decoder architecture was presented in [10], where the authors propose use of the convolutive transfer function (CTF) to approximate a time domain convolution. The neural network is trained to predict a channel shortening filter such that, when it is convolved with the STFT of the input signal, the result is a dereverberated signal retaining only the early reflections. In this work, we implemented this architecture and increased the frequency resolution of the STFT to 1024 as this was found to improve performance. We also changed the shortening target to 50 ms to make it comparable with the WPE method. The CS model was trained with speech from the Wall Street Journal (WSJ) database [14], augmented with simulated RIRs covering the range of T60 values from 200 – 600 ms followed by the addition of white and ambient noise in the 5 – 25 dB SNR range.

### 2.3. Fixed MVDR Beamformer (FMB)

This is a time-invariant filter-and-sum beamformer designed according to the Minimum Variance Distortionless Response criteria [3]. The filter weights, $\mathbf{w}$ for particular look direction are computed to minimize the output noise power from the beamformer (minimum variance) subject to constraining the gain in the look direction to be unity (distortionless response), as follows.

$$\mathbf{w} = \frac{\mathbf{v}^{\mathsf{H}}(\mathbf{\Phi}_N + \sigma_d^2 \mathbf{I})^{-1}}{\mathbf{v}^{\mathsf{H}}(\mathbf{\Phi}_N + \sigma_d^2 \mathbf{I})^{-1} \mathbf{v}}, \quad (1)$$

where $\mathbf{v}$ is a steering vector, $\mathbf{\Phi}_N$ is the noise covariance matrix and $\sigma_d^2 \mathbf{I}$ represents the diagonal loading of the noise co-variance matrix for increased robustness [3]. In this paper, we design two types of FMB beamformers, one based on a spatially uncorrelated and another on a spherically isotropic noise field. In the case of the FMBu beamformer, we assume that the noise on each sensor is uncorrelated noise ($\boldsymbol{\Phi}_N = \mathbf{I}$) and this configuration is equivalent to a delay-and-sum beamformer. In the case of the FMBi beamformer design, we assume that the noise field is spherically isotropic where equal noise energy is received from all directions around the array. In this case, we set the noise covariance matrix to be $\boldsymbol{\Phi}_N = \text{sinc}(2\pi f d/c)$, where $\text{sinc}\, x = \sin x/x$, $d$ is the inter-microphone spacing and $c$ is the speed of sound. This solution is known as a superdirective beamformer.

### 2.4. Self-Attention Channel Combinator (SACC)

This front-end is introduced in [11]. It produces a single-channel STFT power spectrogram $\mathbf{Y} \in \mathbb{R}^{T \times F}$ by taking the element-wise product and sum over the channel dimension of the weights matrix $\mathbf{W} \in \mathbb{R}^{T \times M \times 1}$ and the normalized logarithmic power of multi-channel input $\mathbf{X} \in \mathbb{R}^{T \times M \times F}$, $\mathbf{Y} = \mathbf{W}\mathbf{X}^\mathsf{T}$, where $T$ and $F$ are the numbers of time frames and frequency bins, respectively. A scaled dot-product self-attention mechanism is utilized to compute the weights $\mathbf{W}$. Three dense layers with linear activations transform $\mathbf{X}$ into the query $\mathbf{Q} \in \mathbb{R}^{T \times M \times K}$, key $\mathbf{K} \in \mathbb{R}^{T \times M \times K}$ and value $\mathbf{V} \in \mathbb{R}^{T \times M \times 1}$, where $K$ is the dimension of the linear transform for producing the query and key. Since we need the weights that work homogeneously on all frequency bins, we use a single unit dense layer to contract the corresponding dimension of $\mathbf{X}$ and to produce the value $\mathbf{V}$. The self-attention matrix $\mathbf{A} \in \mathbb{R}^{T \times M \times M}$ is given by $\mathbf{A} = \text{softmax}\left(\frac{\mathbf{QK}^\mathsf{T}}{\sqrt{D}}\right)$. The weights matrix is calculated as $\mathbf{W} = \text{softmax}(\mathbf{AV}^\mathsf{T})$. In both cases, the softmax is applied to the last channel dimension of the product.

### 2.5. Neural Spatial Filtering (NSF)

The NSF beamformer [4] is a complex-valued neural network layer that models a fixed (time-invariant) filter-and-sum beamformer design, with the advantage of optimizing the WER (instead of the typical SNR based signal criteria). A set of 8-channel filters are learnt during training and fixed during testing. In this paper, we use 16 beams (look directions) for filtering the array signal. The filtered signal energy in 16 directions are then combined to form a single channel representation. The NBF takes all array channels as the input and is jointly optimized with the ASR backend.

### 2.6. Adaptive MVDR Beamformer (AMB)

This is an adaptive MVDR beamformer that uses a spectral mask calculated from the coherent-to-diffuse ratio [15] to better estimate the speech covariance matrix in a diffuse noise field. More details can be found in [11].

Table 1. ASR results for different system configurations.

| Method | Word Error Rate (WER, %) | | | | | |
|---|---|---|---|---|---|---|
| | P1 | P2 | P3 | P4 | Avg. | Rel. |
| SDM | 10.4 | 10.3 | 11.7 | 10.3 | 10.7 | - |
| AMB | 9.7 | 8.5 | 9.9 | 9.0 | 9.3 | 13.1 |
| NSF | 8.9 | 8.8 | 9.6 | 8.4 | 8.9 | 16.8 |
| SACC | 9.1 | 8.4 | 9.1 | 8.7 | 8.8 | 17.8 |
| SACC + | | | | | | |
| CS | 9.3 | 8.5 | 9.5 | 9.1 | 9.1 | 15.0 |
| WPEs | 9.5 | 8.7 | 9.6 | 8.5 | 9.1 | 15.0 |
| WPEm | 7.6 | 7.1 | 7.8 | 6.9 | 7.4 | 30.8 |
| FMBu | 8.9 | 8.8 | 9.7 | 8.5 | 9.0 | 15.9 |
| FMBi | 10.0 | 9.6 | 9.9 | 9.1 | 9.7 | 9.3 |
| WPEm-NSF | 7.0 | 7.0 | 7.7 | 6.7 | 7.1 | 33.6 |
| **WPEm-FMBu** | **7.0** | **6.7** | **7.5** | **6.5** | **6.9** | **35.5** |

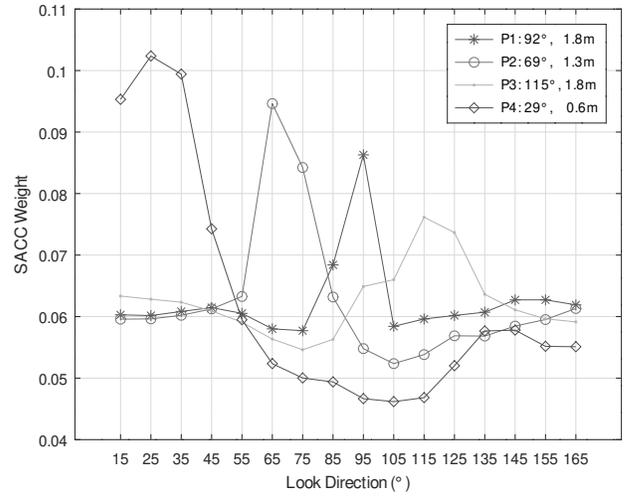

**Fig. 2**. Average SACC weights for test set from each playback position with oracle azimuth and range labeled in the legend).

## 3. EVALUATION AND RESULTS

### 3.1. Data

In order to train the ASR systems, multi-channel training data was created by convolving a clean subset of single channel speech from LibriSpeech [17] and English partition of the Mozilla Common Voice[1] with simulated room impulse responses (RIRs). Nearly 460 hr of speech was selected by placing thresholds on signal parameters estimated by NISA [11, 16]. This speech material was convolved with simulated RIRs for an 8 channel uniform linear array with 33 mm inter-mic. spacing and a number of directional sources placed at random positions in a large set of rooms (T60 in range $0.3 - 0.8$ s). Finally, ambient noise was added to the material in an SNR range of 5 to 25 dB and 45 dB SNR white noise simulating microphone self-noise. The level of each microphone channel within an array was also augmented in the 0.1 to 2 dB range and finally, the overall signal level was augmented in the −1 to −15 dBFS range. For test data, a set of clean speech utterances from the Libri-test-clean partition were simultaneously played back through artificial mouth loudspeakers and recorded by a wall mounted 8 microphone uniform linear array. The playback and recording was performed from four positions in a typical office room (3 m × 3.7 m), simulating two pairs of conversation positions. The speaker pairs were oriented toward each other. Oracle azimuth and distance for each position are shown in Fig. 2.

---
[1]https://commonvoice.mozilla.org/en

**Table 2**. Estimated C50 (dB) using NISA [16] for some of the front-end systems. NISA has a 3.3 dB mean absolute error for C50 estimation.

| Method | C50 (dB) | | | | |
|--------|------|------|------|------|------|
|        | P1   | P2   | P3   | P4   | Avg. |
| SDM    | 8.5  | 8.4  | 7.2  | 8.8  | 8.2  |
| AMB    | 16.3 | 15.8 | 15.6 | 16.6 | 16.1 |
| CS     | 15.6 | 16.4 | 16.1 | 16.4 | 16.1 |
| WPEs   | 5.5  | 7.1  | 8.3  | 8.8  | 7.4  |
| WPEm   | **21.5** | **21.8** | **21.8** | **21.9** | **21.8** |

### 3.2. Experiments

We experiment with WPE and CS based pre-processing followed by an optional FMB beamformer and then a SACC-based ASR front-end. We also investigate the combination of WPE with the learn-able NSF beamformer and the SACC front-end. The number of fixed beams for both FMB and NSF are set to 16. An attention-based encoder-decoder (AED) E2E ASR system for all experiments, with the encoder based on ContextNet [18] and a single layer LSTM decoder [11]. For all experiments, the ASR system in trained for 90 epochs. The ASR performance with different front-ends is evaluated in terms of the word error rate (WER). A baseline condition of single distant microphone (SDM) is also included for reference (this is an ASR model trained with channel 4 of the simulated training data and tested also on channel 4 of the test set).

### 3.3. Results

The ASR results are shown in Table 1, where we can see the WER for the four playback positions (P1 to P4). The baseline SDM system achieves an average WER of 10.7% and is outperformed by all other systems tested. In a previous paper, we had shown how the SACC based system with raw microphone input gave the best WER [11] and here too it outperforms both the AMB and NSF methods.

The pre-processing of the raw input signals to SACC by the single channel CS and SISO WPE (WPEs) methods degrade the overall WER as compared to SACC with raw data. This may be attributed to insufficient dereverberation performed with single channel methods. In Table 2 we estimate the C50 metric using NISA [16] and observe that the WPEs method does not significantly change the C50 from the SDM condition and that the overall C50 estimates correlate with the WER results presented in Table 1. We can see that the MIMO WPEm method based pre-processing has the best improvement in C50 and when used with SACC for ASR, this results in a large improvement over the SACC baseline.

On their own, the fixed MVDR based beamforming front-end do not outperform the baseline SACC system, and we see that the FMBu design performs better than FMBi. The best overall result is obtained with the combination of WPEm and FMBu front end to SACC. This front-end provides a 35.5% WER reduction relative to the SDM model and a 21.6% WER reduction relative to the SACC model with raw audio input. It is noteworthy that this combination outperforms also the NSF based combination with WPEm and SACC, which achieves the second lowest WER of 7.1%.

An explanation for the WPEm-FMBu system working well is that the FMBu design is based on ideal steering vectors (assuming anaechoic conditions) and thus performs well only when the reverberation is low. This design results in clear beams unlike the NSF based system, as shown in Fig. 1, the beampattern for the FMBu system for a 95° look direction has a clear beam with 0 dB gain in the look direction whereas the NSF based filters that are learnt from the training data do not have a clear beam direction. Thus, a further advantage of the proposed WPEm-FMBu system is that it allows for spatial localization to be performed by investigating the SACC weights estimated on a test utterance. In Fig. 2 we show the average SACC weights for the 16 fixed FMBu beams for the four test positions. We can clearly see that the peak at each position corresponds to the beam that is closest to the actual azimuth of the playback setup. Moreover, the flatness of the SACC weight profile provides information on the diffuseness of the sound field (i.e. in an anechoic room we would expect to see impulses). This information has many potential applications including speaker diarization and localization, which is something we will explore in future work.

### 4. CONCLUSIONS

We presented a new acoustic front-end for multi-channel E2E ASR based on channel shortening with a MIMO WPE method followed by a fixed MVDR beamformer and a SACC system for beam selection and combination. Our proposed approach outperforms our previous SACC system by 21.5% in terms of relative WER. Moreover, we show how the SACC weights provide accurate spatial information that could be used in future work as features for speaker diarization.

# 5. REFERENCES

[1] R. Haeb-Umbach, J. Heymann, L. Drude, S. Watanabe, M. Delcroix, and T. Nakatani, "Far-Field Automatic Speech Recognition," *Proceedings of the IEEE*, vol. 109, no. 2, pp. 124–148, Feb. 2021.

[2] T. Ochiai, S. Watanabe, T. Hori, and J. R. Hershey, "Multichannel End-to-end Speech Recognition," in *International Conference on Machine Learning*. July 2017, PMLR.

[3] K. Kumatani, T. Arakawa, K. Yamamoto, J. McDonough, B. Raj, R. Singh, and I. Tashev, "Microphone array processing for distant speech recognition: Towards real-world deployment," in *Proc. of The 2012 Asia Pacific Signal and Information Processing Association Annual Summit and Conference*, 2012, pp. 1–10.

[4] W. Minhua, K. Kumatani, S. Sundaram, N. Strom, and B. Hoffmeister, "Frequency Domain Multi-channel Acoustic Modeling for Distant Speech Recognition," in *Proc. of ICASSP*, May 2019.

[5] C. Boeddeker, H. Erdogan, T. Yoshioka, and R. Haeb-Umbach, "Exploring Practical Aspects of Neural Mask-Based Beamforming for Far-Field Speech Recognition," in *Proc. of ICASSP*, Apr. 2018.

[6] J. Heymann, L. Drude, and R. Haeb-Umbach, "Neural network based spectral mask estimation for acoustic beamforming," in *Proc. of ICASSP*. 2016, p. 196–200, IEEE Press.

[7] T. Ochiai, S. Watanabe, T. Hori, J. R. Hershey, and X. Xiao, "Unified architecture for multichannel end-to-end speech recognition with neural beamforming," *IEEE Journal of Selected Topics in Signal Processing*, vol. 11, no. 8, pp. 1274–1288, 2017.

[8] B. Li, T. N. Sainath, R. J. Weiss, K. W. Wilson, and M. Bacchiani, "Neural Network Adaptive Beamforming for Robust Multichannel Speech Recognition," in *Proc. of Interspeech*, 2016.

[9] T. Nakatani, T. Yoshioka, K. Kinoshita, M. Miyoshi, and B.-H. Juang, "Speech dereverberation based on variance-normalized delayed linear prediction," *IEEE Transactions on Audio, Speech, and Language Processing (TASLP)*, vol. 18, no. 7, pp. 1717–1731, 2010.

[10] H. Chung, V. S. Tomar, and B. Champagne, "Deep convolutional neural network-based inverse filtering approach for speech de-reverberation," in *2020 IEEE 30th International Workshop on Machine Learning for Signal Processing (MLSP)*, 2020, pp. 1–6.

[11] R. Gong, C. Quillen, D. Sharma, A. Goderre, J. Lainez, and L. Milanovic, "Self-attention channel combinator frontend for end-to-end multichannel far-field speech recognition," in *Proc. of Interspeech*, 2021.

[12] T. Yoshioka and T. Nakatani, "Generalization of multichannel linear prediction methods for blind mimo impulse response shortening," *IEEE TASLP*, vol. 20, no. 10, pp. 2707–2720, 2012.

[13] L. Drude, J. Heymann, C. Boeddeker, and R. Haeb-Umbach, "Nara-wpe: A python package for weighted prediction error dereverberation in numpy and tensorflow for online and offline processing," in *Speech Communication; 13th ITG-Symposium*, 2018, pp. 1–5.

[14] CSR-II (WSJ1), "Complete LDC94S13A. DVD. Philadelphia: Linguistic Data Consortium, 1994.," .

[15] A. Schwarz and W. Kellermann, "Coherent-to-diffuse power ratio estimation for dereverberation," *IEEE TASLP*, vol. 23, no. 6, pp. 1006–1018, 2015.

[16] D. Sharma, L. Berger, C. Quillen, and P. A. Naylor, "Non intrusive estimation of speech signal parameters using a frame based machine learning approach," in *Proc. of EUSIPCO*, Amsterdam, The Netherlands, 2020.

[17] V. Panayotov, G. Chen, D. Povey, and S. Khudanpur, "LibriSpeech: An ASR corpus based on public domain audio books," in *Proc. of ICASSP*, Brisbane, Australia, 2015, IEEE.

[18] W. Han, Z. Zhang, Y. Zhang, J. Yu, C.-C. Chiu, J. Qin, A. Gulati, R. Pang, and Y. Wu, "ContextNet: Improving Convolutional Neural Networks for Automatic Speech Recognition with Global Context," in *Interspeech 2020*, Oct. 2020.